# Calibration of the oscillation amplitude of electrically excited scanning probe microscopy sensors

Omur E. Dagdeviren[1,*], Yoichi Miyahara[1], Aaron Mascaro[1], Peter Grütter[1]

[1] Department of Physics, McGill University, Montréal, Québec, Canada, H3A 2T8
*Corresponding author's email: omur.dagdeviren@mcgill.ca

Atomic force microscopy (AFM) is an analytical surface characterization tool which can reveal a sample's topography with high spatial resolution while simultaneously probing tip-sample interactions. Local measurement of chemical properties with high-resolution has gained much popularity in recent years with advances in dynamic AFM methodologies. A calibration factor is required to convert the electrical readout to a mechanical oscillation amplitude in order to extract quantitative information about the surface. We propose a new calibration technique for the oscillation amplitude of electrically driven probes, which is based on measuring the electrical energy input to maintain the oscillation amplitude constant. We demonstrate the application of the new technique with quartz tuning fork including the qPlus configuration, while the same principle can be applied to other piezoelectric resonators such as length extension resonators, or piezoelectric cantilevers. The calibration factor obtained by this technique is found to be in agreement with using thermal noise spectrum method for capsulated, decapsulated tuning forks and tuning forks in the qPlus configuration.

Dynamic scanning probe microscopy is a surface characterization technique where a sharp tip is attached to the end of an oscillating probe and acts as a sensing element to map the surface topography up to picometer resolution.[1-4] With the advent of dynamic scanning probe techniques, quantitative measurement of local sample properties became popular.[5-8] To extract quantitative information about the surface and to modulate the tip-sample separation in a controlled way, properties of the oscillating probe such as excitation frequency, excitation amplitude, frequency shift due to tip-sample interaction, phase difference between the excitation signal and the oscillation signal, quality factor, and oscillation amplitude have to be known.[9-13] While the excitation frequency and the amplitude are directly controlled by the operator, the resonance frequency shift, phase difference, and quality factor are measured by well-established measurement electronics.[2,14] Among these dynamic properties, the oscillation amplitude requires determination of a calibration factor between the electrical readout and the mechanical oscillation amplitude. The knowledge of the oscillation amplitude is important to quantitatively and even qualitatively interpret frequency shift measurements ('small amplitude' vs. 'large amplitude' approximations).[9-13] Accurate knowledge of the oscillation amplitude is key to determine the tip-sample interactions laws from frequency shift data and, as recently demonstrated, needs to be known to determine if it is even in principle possible, i.e. a mathematically well-posed problem.[15]

Many techniques exist to determine the oscillation amplitude, including thermal excitation,[16,17] interferometric techniques,[18,19] electro-mechanical techniques,[20] and frequency modulation-based techniques[16,21]. These existing techniques have limitations when applied to tuning forks and tuning fork-based sensors: Measuring thermal excitations of tuning forks at low temperatures are limited by the microscope's mechanical and electrical noise detection.[17] Specifically, the high $Q$ factors (leading to mechanical noise sensitivity) and high spring constant (resulting in small amplitudes) impede measurement of the thermal noise spectrum of tuning fork system at low-temperatures.[17] Interferometric measurements can only be implemented with the integration of complex optical setups with microscopes that work with tuning fork-based sensors.[18,19] Existing electro-mechanical techniques can only be applied to balanced tuning forks.[20,22] Frequency modulation-based techniques rely on measuring the frequency shift induced by tip-sample interaction upon approach allowing the calibration of the oscillation amplitude due to the indirect measurement of current and frequency shift dependencies.[21] In passing we note that frequency shift-based methodologies can only be conducted with assembled sensors, and in particular oscillation amplitude determination based on tunneling currents requires a conductive sample.[21]

In this manuscript, we propose a new calibration procedure that relies only on measuring the electrical energy input to maintain the oscillation amplitude of the probe constant. We demonstrate the technique with quartz tuning forks and quartz tuning forks in the qPlus configuration and compare our results with the thermal excitation technique at room temperature. The major advantage of our technique is that it directly delivers the calibration factor between the electrical readout and the mechanical oscillation, thereby eliminating the limitations of existing methodologies. As an outlook, we provide a pathway to track the non-negligible change of the spring constant when a tuning fork-based sensor is assembled and mounted. Although the spring constant of tuning forks can be assessed accurately (*see supplemental information for details*), the equivalent spring constant of the sensor assembly can be significantly different compared to tuning forks without tips[23,24] leading to a large systematic error for quantitative force measurements if simply assumed constant[13].

As Figure 1 summarizes, a sinusoidal voltage, $V_{drive}$, is applied to a tuning fork and the resulting oscillating output current, $I_{out}$, is measured. The applied sinusoidal voltage results in a mechanical oscillation of the tuning fork prongs with oscillation amplitude, $A_{osc}$, due to piezoelectric nature of quartz. The mechanical oscillation of the tuning fork's prongs in turn induces an oscillating charge on the surface of the tuning fork, which is collected with electrodes. The time-varying charge is measured as an electric current which is converted by a transimpedance amplifier to voltage and typically demodulated with a lock-in amplifier. The lock-in outputs the phase between the excitation signal and the oscillation signal in degrees ($\varphi$) and oscillation amplitude, $V_{readout}$, in volts.

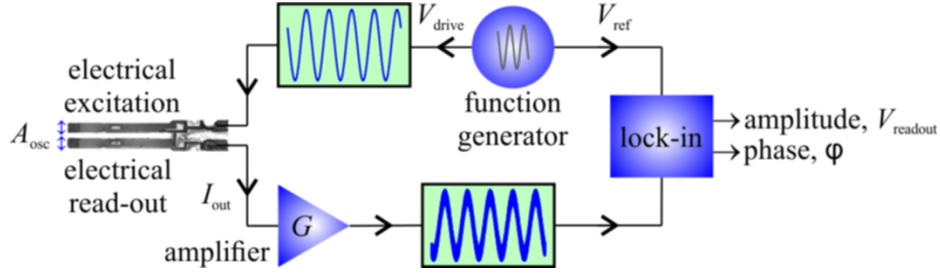

**Figure 1:** Schematic representation of the experimental setup. The piezoelectric probe is excited electrically, $V_{drive}$, which results in a mechanical oscillation, $A_{osc}$. The mechanical oscillation induces an oscillating current, $I_{out}$, which is converted to voltage with the (transimpedance) amplifier. The output of the amplifier is demodulated with the lock-in. The phase difference between the excitation signal and the oscillation signal in degrees ($\varphi$) and oscillation amplitude in volts, $V_{readout}$, are outputs of the lock-in amplifier.

In the absence of a tip-sample interaction, the dynamics of the probe can be expressed as a damped-harmonic oscillator. Equation 1 describes the total energy of a harmonic oscillator:[25]

$$E_{mechanical} = \frac{1}{2}kA_{osc}^2 \qquad (1)$$

In equation 1, $A_{osc}$ is the mechanical oscillation amplitude and $k$ is the equivalent spring constant of the probe. We calculated the spring constant by using finite element methods (*see supplemental information for details*). The total mechanical energy of a balanced tuning fork is equal to $kA_{osc}^2$ as there are two oscillating prongs.[17,22]

Due to the finite quality factor, $Q$, of oscillating probes, the energy dissipated per oscillation cycle is represented by:

$$E_{diss} = \frac{2\pi}{Q}E_{mechanical} \qquad (2)$$

The energy dissipated per second can be calculated by $E_{diss} \times f_0$, where $f_0$ is the resonance frequency of the oscillating probe. To keep the oscillation amplitude constant, this mechanical power loss is compensated by the electrical power, $P_{compensation}$, supplied by the external circuit such as $E_{diss} \times f_0 = P_{compensation}$. Equation 3 shows the average electrical power input to compensate dissipated energy:

$$P_{compensation} = \frac{1}{2}V_{drive}I_m \cos\varphi \qquad (3)$$

In equation 3, $V_{drive}$ is the amplitude of the sinusoidal excitation voltage and $I_m$ is the current induced by the mechanical oscillation and $\varphi$ is the phase of $I_m$ with respect to $V_{drive}$. The phase, $\varphi$, is equal to zero at the resonance frequency, and equation 3 reduces to $\frac{1}{2}V_{drive}I$. The total measured current is equal to $I_{out} = V_{readout}/G$, where $G$ is the gain of the measurement setup (*see supplemental information for details*). It is important to notice that $I_{out}$ is the sum of $I_m$ and the current that passes through the stray capacitance.[26-29] We calculated $I_m$ (= $V_{compensated}/G$), with the analytical formula by Lee et al.[29], alternative correction techniques are available.[26-28] Finally, we can find the oscillation amplitude of a probe by equating the mechanical energy dissipation to electrical energy input to maintain the oscillation amplitude constant:

$$A_{osc} = \sqrt{\frac{V_{drive} \times V_{readout} \times Q}{2\pi \times f_0 \times k \times G}} \tag{4}$$

Equation 4 allows a linear fit of the mechanical amplitude in Ångströms to the electrical signal in volts:

$$A_{osc} = \alpha \times V_{readout} \tag{5}$$

In Equation 5, $\alpha$ is the calibration factor. The calibration of a decapsulated tuning fork with the principle of energy balance is demonstrated in Figure 2. Several resonance curves are measured with different drive amplitudes. As Figure 2a shows, the frequency response of the tuning fork deviates from the ideal Lorentzian shape with an anti-resonance peak due to the effect of stray capacitance. Figure 2b shows the corrected frequency response obtained from results in Figure 2a using methods described in Ref [29]. Finally, the oscillation amplitude, $A_{osc}$, is calculated with Equation 4. Figure 2c shows the linear relation (i.e. the calibration factor, $\alpha$ as presented in equation 5) of a tuning fork probe for different drive signals both with direct readout, i.e. without correction of current through stray capacitance (red curve), and the corrected readout (blue curve). In Equation 4 and 5, we use $V_{compensated}$ instead of $V_{readout}$ for the calibration with the effect of stray capacitance corrected.

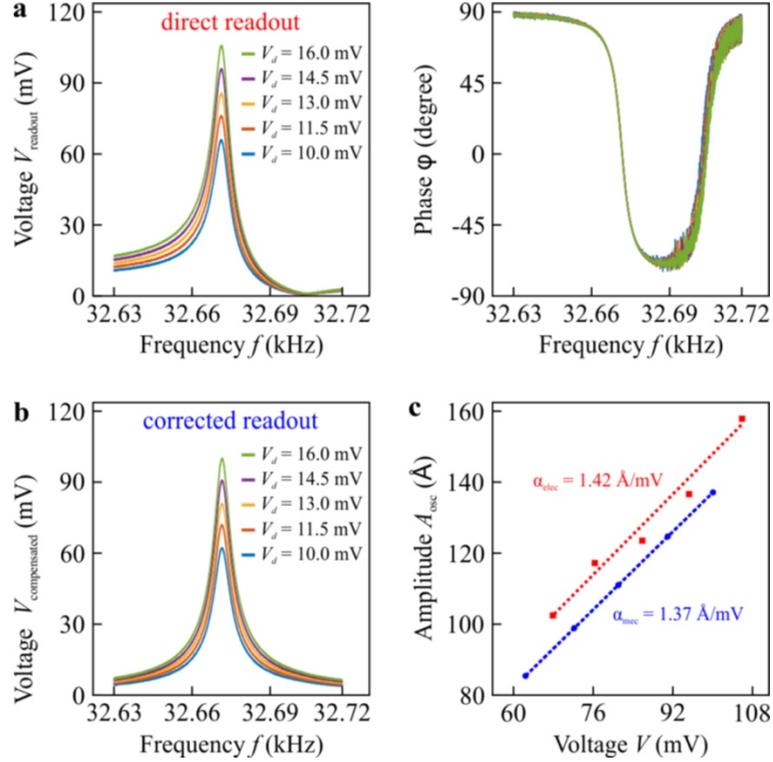

**Figure 2:** Calibration of an electrically excited tuning fork with the principle of energy balance. (a) shows the direct electrical readout of the oscillation amplitude and the phase. (b) The mechanical equivalent of the readout is compensated by using the mathematical correction procedure.[29] (c) The calibration constant of the oscillation amplitude is calculated (blue, corrected readout and red, direct readout). The horizontal axis of the plot in (c) is $V_{readout}$ for calibration with the direct readout and $V_{compensated}$ with the compensation of stray capacitance. The confidence level of fits for calibration factor, $\alpha$, is 1 picometer/mV, which is consistent with former experimental work [16,21] and results in 0.7% uncertainty due to the curve fitting. The tuning fork used for experiments presented in Figure 2 has spring constant of 1,267 N/m, $Q$ = 5,100, and resonance frequency of 32,673 Hz.

To elaborate the calibration of oscillation amplitude with the principle of energy balance, we applied the technique to three different capsulated and decapsulated tuning forks and compared our results with the thermal excitation technique (*see supplemental information for details*). As summarized in Table 1, calibration with direct readout is in agreement with less than 2.0% discrepancy compared to calibration using the corrected readout and thermal excitation for capsulated tuning forks. However, the calibration factor with direct readout can deviate compared to decapsulated tuning forks. Moreover, this discrepancy upsurges when the quality factor decreases upon decapsulation for all tuning forks and reaches at least 9% for type II tuning fork. We conclude that stray capacitance effects should be corrected to ensure accurate calibration.

|  |  | **Direct Readout** | **Corrected Readout** | **Thermal Excitation** |
|---|---|---|---|---|
| **Type I**, $k$ = 1,267 N/m | *Capsulated, Q = 61,200* | 1.38 Å/mV | 1.38 Å/mV | 1.37 Å/mV |
|  | *Decapsulated, Q = 5,100* | 1.42 Å/mV | 1.37 Å/mV | 1.37 Å/mV |
| **Type II**, $k$ = 1,939 N/m | *Capsulated, Q = 71,900* | 1.10 Å/mV | 1.08 Å/mV | 1.09 Å/mV |
|  | *Decapsulated, Q = 5,800* | 1.16 Å/mV | 1.06 Å/mV | 1.06 Å/mV |
| **Type III**, $k$ = 16,940 N/m | *Capsulated, Q = 101,000* | 0.29 Å/mV | 0.28 Å/mV | 0.28 Å/mV |
|  | *Decapsulated, Q = 9,500* | 0.29 Å/mV | 0.28 Å/mV | 0.28 Å/mV |

**Table 1:** Calibration of oscillation amplitude for three different types of capsulated and decapsulated tuning forks with direct readout, corrected readout, and thermal excitation technique.

Quartz tuning forks that have one free prong to which the tip is attached to the end while the fork's other prong is fixed to a holder ('qPlus' configuration) have gained popularity in recent years for high-resolution imaging.[30] Figure 3a schematically describes that when a tuning fork in qPlus configuration is excited electrically, the electrical excitation voltage is applied to both prongs; however, only one of the prongs is free to oscillate (i.e. the effective impedance at resonance is half of the tuning fork configuration). Figure 3b shows the amplitude and phase of the tuning fork in qPlus configuration with direct electrical readout. We corrected the electrical readout to eliminate the effect of stray capacitance (Figure 3c). The calibration factor with the corrected readout, $\alpha_{mec}$ = 2.70 Å/mV (Figure 3d, blue curve) is in agreement with the calibration of the same tuning fork in qPlus configuration with thermal excitation, $\alpha_{thermal}$ = 2.72 Å/mV, the 1% difference well within the uncertainty of the thermal excitation technique.[16,17,31] Similar to the previous observation, the calibration factor determined from a direct readout (Figure 3d, red curve) deviates by 15% with respect to the calibration with thermal excitation. Such a drastic difference emphasizes the importance of correcting the stray capacitance of tuning forks for accurate calibration of oscillation amplitude when using the principle of energy balance.

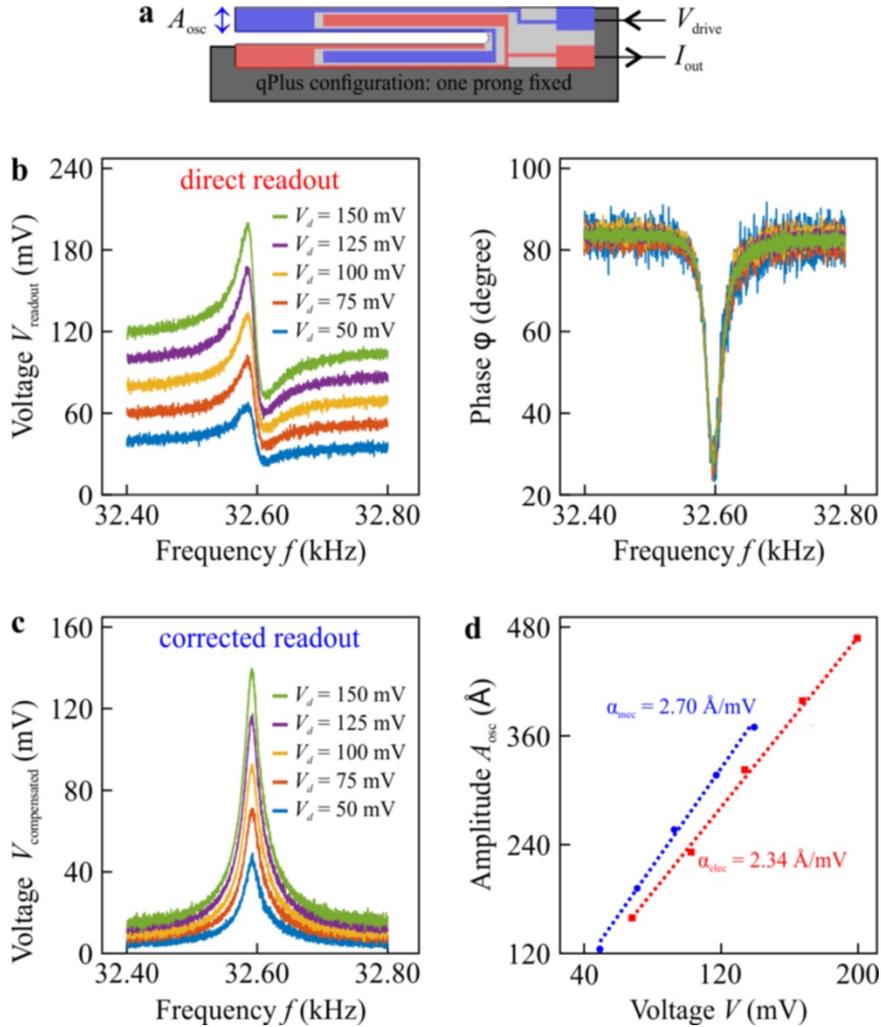

**Figure 3:** Calibration of an electrically excited tuning fork in qPlus configuration with the principle of energy balance. (a) One of the prongs of the tuning fork is fixed while the other prong oscillates freely for tuning forks in qPlus configuration. (b) shows the direct electrical readout of the electrically excited tuning fork in the qPlus configuration for different drive amplitudes. (c) The effect of stray capacitance is corrected with the mathematical correction procedure.[29] (d) The calibration constant of the oscillation amplitude is calculated both with direct readout (red curve) and corrected readout (blue curve) by curve fitting to the experimental data. The confidence level of fits for calibration factor, $\alpha$, is 1 picometer/mV, which is consistent with former experimental work [16,21] and results in 0.7% uncertainty due to the curve fitting. The horizontal axis of the plot in (d) is $V_{readout}$ for calibration with the direct readout and $V_{compensated}$ with the compensation of stray capacitance. For experiments presented in Figure 3, we used type I tuning fork in qPlus configuration, $Q = 1,416$, and resonance frequency of 32,592 Hz.

In the previous sections we have only compared tuning forks as fabricated, without any tip or tip-wires attached. Although the spring constant of the tuning forks can be quantified accurately with experimental and computational techniques (*see supplemental information for details*) [21,30,32-34], the spring constant of the tip-tuning fork assembly is different than the tuning fork alone. Knowing the spring constant of the assembled tip-tuning fork system is particularly important for quantifying interaction potentials from AFM measurements and is normally not easily accessible[21,24,32,33]. The spring constant of the sensor assembly is expected to be potentially substantially different compared to the bare tuning fork as the mechanical constraints, i.e. *boundary conditions*, change due to the epoxy glue that is used to attach the tuning fork to the base, the epoxy that is used to attach the tip, the orientation of the tip, and the wire to collect tunneling current or apply a tip-bias in AFM.[21,24,32,33] In the following, we will show how a measurable change in the calibration factor, $\alpha$, can be used to track the change in the spring constant as the sensor is assembled. The calibration factor, $\alpha$, scales as (*see supplemental information for details*):

$$\alpha \propto 1/(f_0 \times k \times Number\ of\ Oscillating\ Prongs) \qquad (6)$$

Since the change in $f_0$ can easily be traced by a frequency sweep, the change in $\alpha$ can be used to quantify the change in spring constant of the sensor, in particular in the qPlus configuration. As an example, the calibration factor, $\alpha$, of a qPlus sensor (2.70 Å/mV, Figure 3c) has a 2% difference compared to the calibration of the same sensor in the tuning fork configuration (Table 1, 1.37 Å/mV × 2 = 2.74 Å/mV) if the same effective spring constant is assumed and the number of oscillating prongs is taken into account (equation 6). The resonance frequency of the sensor in the qPlus configuration has only a 0.1% drop (Figure 3) with respect to the decapsulated tuning fork (Figure 2). According to Equation 6, such a small change in resonance frequency has a negligible effect on $\alpha$ and thus the proportionality presented in equation 6 reduces to $\alpha \propto 1/k$. For this reason, the 2% decrease in the calibration factor implies the effective spring constant of the qPlus sensor as 1,293 N/m. Similarly, the calibration factor of type II tuning fork in the qPlus configuration is found as 1.97 Å/mV with the principle of energy balance, which reveals the effective spring constant as 2087 N/m. The reassessed spring constant of type II tuning fork 4% higher than the generally accepted value (2000 N/m) for similar qPlus sensors.[21,32-34] Note, however, the variation of the spring constant can be more dramatic due to the orientation of the tip and the wire to collect tunneling current or apply a tip-bias in AFM.[23,24] With the reassessment of the spring constant, a systematic error when quantitatively reconstructing the tip-sample interaction laws can be avoided.[13,35]

In summary, we demonstrate a new calibration technique for the oscillation amplitude of electrically driven piezoelectric probes using the principle of energy balance. We demonstrated the application of this new calibration technique with tuning forks including the qPlus configuration, while the same principle can be applied to other piezoelectric oscillators such as length extension resonators, or piezoelectric cantilevers. Our experimental results show that the calibration with the principle of energy balance can be applied independently of the quality factor and the sensor configuration, as long as the effect of stray capacitance is compensated. In addition to revealing the conversion factor between the electrical readout and the mechanical oscillation amplitude, our methodology provides a pathway to track the change in effective spring constant of a sensor assembly in qPlus configuration which is important for quantitative force spectroscopy experiments.

Financial support from The Natural Sciences and Engineering Research Council of Canada and Le Fonds de Recherche du Québec - Nature et Technologies are gratefully acknowledged.

# Supplemental Information:

## Calibration of the oscillation amplitude of electrically excited scanning probe microscopy sensors

Omur E. Dagdeviren[1,*], Yoichi Miyahara[1], Aaron Mascaro[1], Peter Grütter[1]

[1] Department of Physics, McGill University, Montréal, Québec, Canada, H3A 2T8
*Corresponding author's email: omur.dagdeviren@mcgill.ca

### I. Calibration of spring constant, $k$

Experimental and theoretical approaches are available to calibrate the spring constant ($k$), i.e. the stiffness, of the probes used for scanning probe microscopy (SPM) experiments (see Ref [1-4] for detailed reviews). We briefly summarize the finite element method (FEM) approach to calculate the spring constant of tuning forks while further details can be found elsewhere.[5-9] All calculations were performed using COMSOL Multiphysics 4.4 structural mechanics software package (COMSOL Multiphysics, GmbH, Berlin-Germany). Modeling the tuning fork for FEM calculations requires measuring the dimensions. We employed calibrated light microscope measurements to obtain the dimensions of the tuning forks (Table SI 1). It is important to reflect the tuning fork's geometry accurately in regions where stress concentrations are expected, e.g. the region between the prongs and where prongs are connected to the base of the tuning fork.[8,9] We implemented a higher mesh density in these regions. We did not include the gold coating, which has an average thickness of 200 Å, nor the notches at the tuning fork base to decrease the cost of computation and modeling. Neglecting these features in our FEM model has been justified as they are mainly important for electrical properties of tuning fork while having no substantial influence on the mechanical properties.[7,8,10]

|  | Width | Length of the Prong | Thickness |
|---|---|---|---|
| **Tuning Fork-I**, $k$ = 1,267 N/m | 234 (μm) | 2471 (μm) | 90 (μm) |
| **Tuning Fork-II**, $k$ = 1,939 N/m | 234 (μm) | 2426 (μm) | 131 (μm) |
| **Tuning Fork-III**, $k$ = 16,940 N/m | 600 (μm) | 3600 (μm) | 250 (μm) |

**Table SI 1:** Dimensions of tuning forks are measured with a calibrated light microscope. Spring constant, $k$, values presented in this table are calculated with finite element method.

In addition to measuring geometric dimensions, assigning relevant material properties such as Young's modulus, Poisson's ratio, and mass density are required. We used the material properties for quartz from the materials library of the FEM software (Table SI 2).

| | Young's modulus | Poisson's ratio | Mass density |
|---|---|---|---|
| **Quartz** | 82 (GPa) | 0.17 | 2648 (kg m$^{-3}$) |

**Table SI 2:** Material properties used for finite element method (FEM) calculations. Properties of quartz from the materials library of the FEM software (COMSOL Multiphysics 4.4) are used.

As Figure SI 1a summarizes, while keeping one of the prongs and the base of the tuning fork rigid, we applied force along $z$-direction to the end of the free prong. We swept the force from 1 μN to 100 μN with 1 μN steps and measured the displacement, $\Delta z$. The next step is fitting a first-order polynomial to find the spring constant of the tuning fork by using Hooke's law. Source code for the FEM model and the spring constant calculations are available as supplemental material.

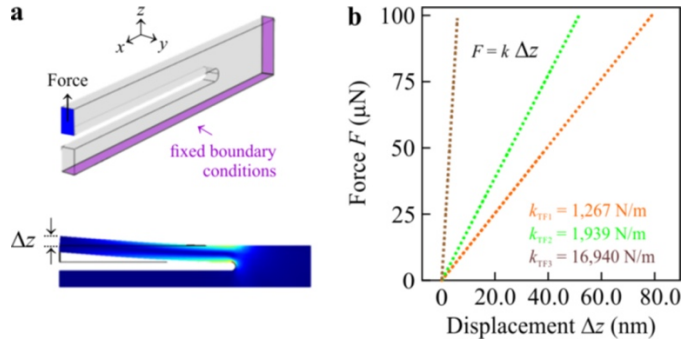

**Figure SI 1:** Calibration of spring constant with finite element methods. (a) One of the prongs and the base of the tuning fork are fixed, while force is applied to the end of the free prong along the $z$-direction. The exerted force deforms the prong and the displacement at the end of the prong, $\Delta z$, is measured. (b) The slope of the force versus displacement curve is equal to the spring constant of the tuning fork, $k$.

Summarized in Figure SI 1b, we find consistent spring constant values with earlier experimental and computational results for similar tuning forks.[5-7,11,12] As outlined by different groups, calibration of spring constant with FEM techniques deviates up to 5% with respect to dynamic experimental results.[5,8,9,13] As the calibration constant between electrical readout and mechanical oscillation depends on the square root of the spring constant (Equation 4, main text), 5% discrepancy in the spring constant will lead to a systematic error of 2.23% for the calibration constant of the oscillation amplitude.

## II. Calibration of gain ($G$) and the phase ($\varphi$)

The time-varying charge induced by the oscillation of the tuning fork prongs are converted to voltage with a current amplifier. As the gain of the amplifiers depends on the frequency,[14,15] the accurate calibration of the gain of the measurement system, $G$, is required to be able to calculate the electrical current induced by the oscillation. The output of the amplifier is demodulated with a lock-in amplifier (Sandford Research Systems, Model SR830 DSP). As Figure SI 2a summarizes, we applied a sinusoidal waveform with the function generator (Agilent, Model 33220A) across an electrical resistor ($R = 1$ G$\Omega$). The equivalent $G$ can be calculated by multiplying the ratio of amplitude readout ($V_{readout}$) and the drive signal ($V_{drive}$) with $R$. As Figure SI 2b reveals, $G$ varies with frequency. We assumed a constant gain as the frequency range of resonance sweep experiments is limited to a few hundred Hz. The inset in Figure SI 2b illustrates that a constant gain of $8.87 \times 10^7$ $\Omega$ leads less than 0.5% standard deviation for 300 Hz window around 32,767 Hz.

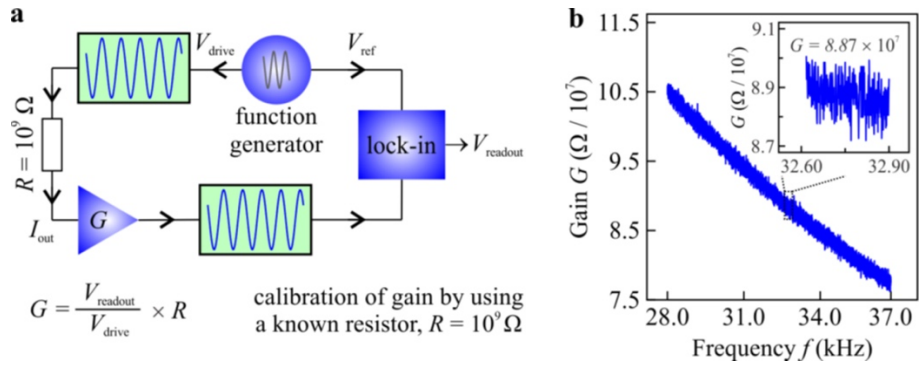

**Figure SI 2:** Calibration of gain of the measurement system, $G$. (a) The gain is equal to the ratio of amplitude readout, $V_{readout}$, and the amplitude of the drive signal, $V_{drive}$, multiplied by the resistor, $R$. (b) The gain changes with the frequency; however, $G$ can be assumed to be constant for small frequency windows.

In addition to calibration of the gain, we calibrated the phase shift due to the measurement electronics and the circuitry. As Figure SI 3 summarizes, we used parallel plate air gap capacitor, which is assumed to be an ideal capacitor with a 90° phase shift between its current and voltage, to calibrate the phase, $\varphi$. Figure SI 3b reveals that the phase (blue curve) evolves with the frequency. We calibrated the phase to 90° at 32,767 Hz and fitted a first order polynomial (red, dashed curve) to calculate the variation of the phase as a function of frequency.

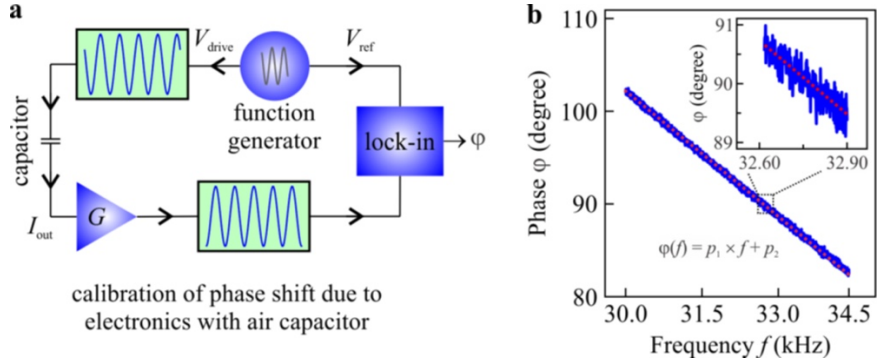

**Figure SI 3:** Calibration of phase shift due to measurement electronics and circuitry. (a) The phase, $\varphi$, is calibrated with an air capacitor. The air capacitor is assumed to be an ideal capacitor with 90° phase shift between its current and voltage. (b) We calibrated the phase shift to 90° at 32,767 Hz and fitted a first order polynomial (red, dashed curve, $p_1 = -0.004326$, $p_2 = 231.6$) to the experimental data (blue curve) to calculate the frequency dependent variation of phase due to measurement electronics and circuitry.

### III. Calibration with thermal excitation

According to the equipartition theorem, thermal noise induces the excitation of the cantilever beam.[16,17] Thermal oscillations of the cantilever can be used for the calibration of oscillation amplitude. We compared results of our technique with results of thermal excitation which has been previously applied to tuning forks and tuning fork-based oscillators.[18-20] Figure SI 4 shows a thermal noise spectrum (blue dots) of a decapsulated tuning fork (type I, $k = 1,267$ N/m) at $T = 295.65$ K. A Lorentzian function (black, dashed curve) with resonance frequency, $f_0 = 32,671.8$ Hz and quality factor, $Q = 5,100$ plus a constant background is fitted to the experimental data which results in a calibration factor of 1.37 Å/mV.

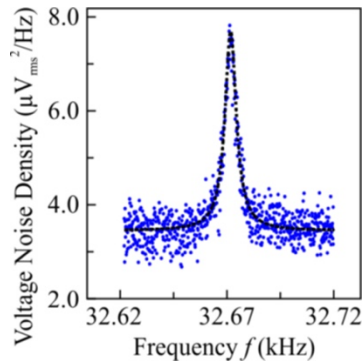

**Figure SI 4:** Thermal noise spectrum of a decapsulated tuning fork. A Lorentzian curve (black, dashed) is fitted to the thermal noise density (blue dots) to calculate the calibration constant of the oscillation amplitude, $\alpha_{thermal} = 1.37$ Å/mV. The spectral noise of the sensor is 255 fm/$\sqrt{Hz}$. The thermal noise density spectrum presented in this figure is averaged 450 times.

## IV. Resonance frequency ($f_0$) and spring constant ($k$) dependence of calibration factor ($\alpha$)

Current induced by mechanical oscillation in the tuning fork is expressed by:

$$I_m = V/Z \qquad \text{(SI 1)}$$

In equation SI 1, $V$ is the amplitude of the electrical voltage across the tuning fork, while $Z$ is the equivalent impedance of the tuning fork.[21] By the relation in equation SI 1, Equation 3 in the main text can be written as:

$$P = \frac{1}{2}\frac{(V)^2}{R} \qquad \text{(SI 2)}$$

In equation 2, $R$ is the equivalent impedance of the tuning fork at the resonance. By inserting equation SI 2 in to the Equation 4 in the main text, the mechanical oscillation amplitude can be expressed as:

$$A_{OSC} = \sqrt{\frac{V \times \frac{V}{R} \times Q}{2\pi \times f_0 \times k}} \qquad \text{(SI 3)}$$

In equation SI 3, $Q$ is the quality factor, while $f_0$ and $k$ are the resonance frequency and the spring constant. With the rearrangement of Equation SI 3, we have the following expression:

$$A_{OSC} = V\sqrt{\frac{Q}{R \times w_0 \times k}} \qquad \text{(SI 4)}$$

In equation SI 4, $w_0$ is the angular frequency ($w_0 = 2\pi \times f_0$). If equation SI 1 and equation SI 4 are introduced in equation 5 in the main text and rearranged for the calibration factor for the current induced due to mechanical oscillation, $\alpha'$:

$$\alpha' = \frac{V\sqrt{\frac{Q}{R \times w_0 \times k}}}{V/R} \qquad \text{(SI 5)}$$

If equation SI 5 is rearranged, we have:

$$\alpha' = \sqrt{\frac{Q \times R}{w_0 \times k}} \qquad \text{(SI 6)}$$

Here, we use the following relation for $R$:[21]

$$R = \frac{L \times w_0}{Q} \qquad \text{(SI 7)}$$

In equation SI 7, $L$ is the equivalent inductance of the tuning fork, and if equation SI 7 is inserted in equation SI 6:

$$\alpha' = \sqrt{\frac{L}{k}} \qquad \text{(SI 8)}$$

The equivalent inductance, $L$, can be written as:[21]

$$L = \frac{m}{\beta^2} \qquad \text{(SI 9)}$$

In equation 9, $m$ is the effective mass of the tuning fork and $\beta$ is the electromechanical coupling constant.[18,19,21] The effective mass can be expressed as:[21]

$$m = \frac{k}{w_0^2} \tag{SI 10}$$

If equation SI 10 is inserted in equation SI 9 and if the resulting equation is inserted in SI 8, the calibration factor per unit current induced due to mechanical oscillation is:

$$\alpha' = \frac{1}{w_0 \times \beta} \tag{SI 11}$$

It has been previously shown that the electromechanical coupling constant, β, is linearly proportional to the spring constant and number of oscillating prongs, i.e. $\beta \propto (k \times Number\ of\ Oscillating\ Prongs)$.[18,19,21] For this reason, the proportionality of the calibration factor can be expressed as:

$$\alpha \propto \frac{1}{w_0 \times k \times (Number\ of\ Oscillating\ Prongs)} \tag{SI 12}$$

**References for Supplemental Information**